\documentclass[sigconf,natbib=true,anonymous=false]{acmart}
\AtBeginDocument{%
  \providecommand\BibTeX{{%
    \normalfont B\kern-0.5em{\scshape i\kern-0.25em b}\kern-0.8em\TeX}}}

\setcopyright{acmlicensed}
\copyrightyear{2024}
\acmYear{2024}
\acmDOI{XXXXXXX.XXXXXXX}

\acmConference[Conference acronym 'XX]{Make sure to enter the correct
  conference title from your rights confirmation emai}{June 03--05,
  2018}{Woodstock, NY}
\acmISBN{978-1-4503-XXXX-X/18/06}

\usepackage{booktabs}
\usepackage{amsmath}
\usepackage{color}
\usepackage{booktabs}
\usepackage{multirow}
\usepackage{makecell}
\usepackage{amsfonts}
\usepackage{mathrsfs}
\usepackage{todonotes}
\usepackage{comment}
\usepackage{multicol}
\usepackage[linesnumbered,boxed]{algorithm2e}
\usepackage{bm}
\usepackage{latexsym}
\usepackage{microtype}
\usepackage[flushleft]{threeparttable}
\begin{document}

\title{Enhancing CTR Prediction through Sequential Recommendation Pre-training: Introducing the SRP4CTR Framework}

\author{Ruidong Han}
\affiliation{%
  \institution{Meituan}
  \city{Beijing}
  \country{China}}
\email[email1]{hanruidong@meituan.com}
\author{Qianzhong Li}
\affiliation{%
  \institution{Meituan}
  \city{Beijing}
  \country{China}}
\email[email1]{liqianzhong@meituan.com}
\author{He Jiang}
\affiliation{%
  \institution{Meituan}
  \city{Beijing}
  \country{China}}
\email[email1]{jianghe06@meituan.com}
\author{Rui Li}
\affiliation{%
  \institution{Meituan}
  \city{Beijing}
  \country{China}}
\email[email1]{lirui84@meituan.com}
\author{Yurou Zhao}
\affiliation{%
  \institution{Gaoling School of Artificial Intelligence, Renmin University of China	}
  \city{Beijing}
  \country{China}}
\email[email1]{zhaoyurou@ruc.edu.cn	}
\author{Xiang Li}
\affiliation{%
  \institution{Meituan}
  \city{Beijing}
  \country{China}}
\email[email1]{lixiang245@meituan.com}
\author{Wei Lin}
\affiliation{%
  \institution{Meituan}
  \city{Beijing}
  \country{China}}
\email[email1]{linwei31@meituan.com}

\renewcommand{\shortauthors}{Ruidong Han, et al.}
\begin{abstract}
Understanding user interests is crucial for Click-Through Rate (CTR) prediction tasks. In sequential recommendation, pre-training from user historical behaviors through self-supervised learning can better comprehend user dynamic preferences, presenting the potential for direct integration with CTR tasks. Previous methods have integrated pre-trained models into downstream tasks with the sole purpose of extracting semantic information or well-represented user features, which are then incorporated as new features. However, these approaches tend to ignore the additional inference costs to the downstream tasks, and they do not consider how to transfer the effective information from the pre-trained models for specific estimated items in CTR prediction. In this paper, we propose a Sequential Recommendation Pre-training framework for CTR prediction (SRP4CTR) to tackle the above problems.
Initially, we discuss the impact of introducing pre-trained models on inference costs. Subsequently, we introduced a pre-trained method to encode sequence side information concurrently.
During the fine-tuning process, we incorporate a cross-attention block to establish a bridge between estimated items and the pre-trained model at a low cost.
Moreover, we develop a querying transformer technique to facilitate the knowledge transfer from the pre-trained model to industrial CTR models. Offline and online experiments show that our method outperforms previous baseline models.
\end{abstract}




\received{20 February 2007}
\received[revised]{12 March 2009}
\received[accepted]{5 June 2009}

\maketitle

\section{Introduction}
Click-through Rate (CTR) prediction is crucial in recommender systems.
Meanwhile, modeling users’ behavior by their sequences is the main approach for CTR prediction, and longer sequences have become a trend in industry recommendation systems\cite{pi2020search, chang2023twin}.



Traditional methods tend to utilize click labels for supervised learning, ignoring the extensive information within the sequence. This information contains not only item IDs but also an abundance of side information, such as price and behavior type. To utilize and model the sequence information, some methods introduce self-supervised pre-training to solve sequential recommendation tasks, whose goals are next-item prediction. 
They have directly demonstrated the potential of integrating self-supervised pre-training with downstream tasks in the recommender system. With this inspiration, we introduce the self-supervised pre-training method for the CTR prediction task. However, self-supervised pre-training methods \cite{sun2019bert4rec, de2021transformers4rec,liu2021noninvasive} for sequence modeling only take item IDs as input or output, and they are insufficient for encoding side information. In fact, fully exploring information in item IDs and side information contributes to the improvement of CTR prediction.

\begin{figure}[!t]
\centering
	\includegraphics[width=1.00\linewidth]{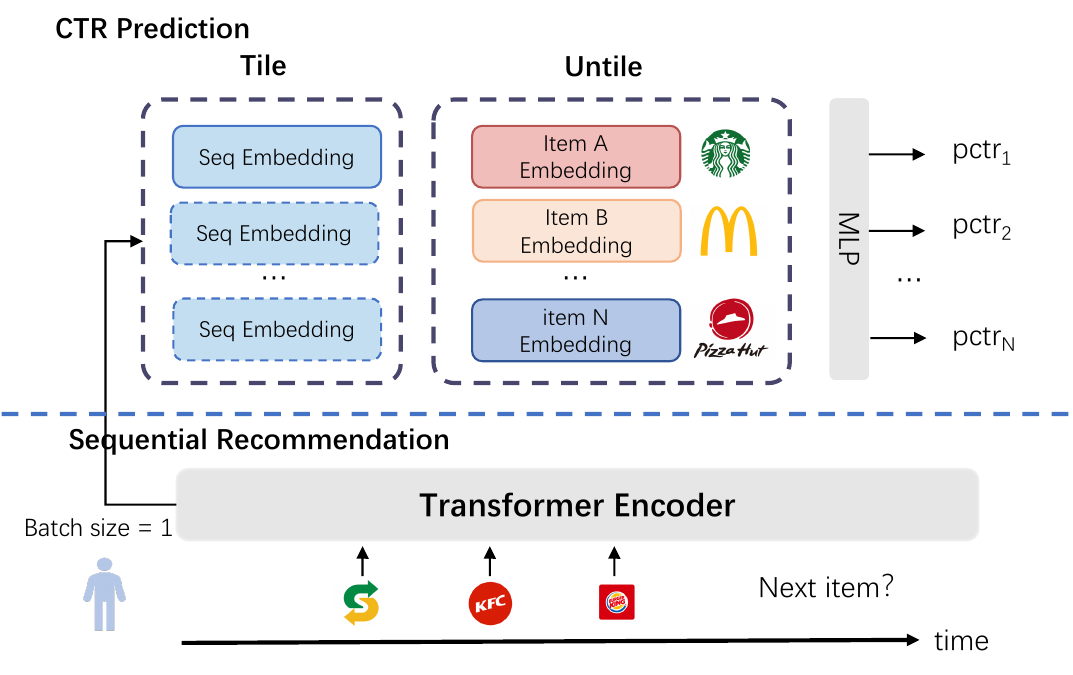}
	\vspace{-0.5em}
	\captionsetup{font={footnotesize}}
	
	\caption{The folded inference pipeline of online deployment. The user behavior encoder can infer a sequential representation with a batch size of 1. Then, the resulting representation can then be replicated multiple times and concatenated with other item features and infer with a batch size of N.}
	\vspace{-1.0em}
    \label{fig:srp}
   
\end{figure}

Another challenge lies in how to integrate pre-training with CTR prediction tasks.
Current methods attempt to integrate CTR prediction with the pre-trained model, which can be categorized based on the form of encoded information as follows: 1) user-related only\cite{liao2020effectiveness,chitlangia2023scaling,liu2022boosting}. 2) user-item-related\cite{lin2023map, wang2023bert4ctr}.
Existing methods have not systematically compared the extra inference costs among different methods, leading to unfair comparison. 
In industrial-level CTR prediction models with extremely high Query Per Second (QPS), the online system utilizes a separate ending workflow for user and item.
For user-related pre-training, as shown in Figure ~\ref{fig:srp}, the user (sequence) information is encoded once, and then this information is tiled across the item, significantly reducing inference costs. In the following paper, we refer to this inference method as folded inference. 
Some approaches\cite{zhai2024actions} have incorporated this concept into the training process. When there is a relatively large number of samples n available for computation under the same user, the $\mathcal{O}(L^2)$ complexity introduced by the transformer can be significantly reduced to a negligible level $\mathcal{O}(L^2/n)$. 
However, the user-related pre-trained model can only obtain a single expression of sequence information, which is then concatenated with item information for learning user interest. These methods fail to capture interest information tailored to the predicted item from the pre-trained model. 
On the contrary, the user-item-related encoding can fully capture the user's interest in the predicted item. However, it causes a great workload for real-time serving and may lead to inapplicability in real scenarios.

In this paper, we propose a new fine-tuning framework designed to adapt the transformer-based Sequential Recommendation Pre-trained model for the CTR prediction task, which we call SRP4CTR.
For encoding item IDs and side information simultaneously, during the pre-training phase, we introduce a new bidirectional transformer model which is named Fine-Grained BERT (FG-BERT). FG-BERT uses all side information in input and output at the same time and performs multi-attribute masking predictions while encoding side information.
For integrating the pre-trained model, in the fine-tuning phase, we introduce a uni cross-attention mechanism that establishes a unidirectional attention connection between the predicted item and the pre-trained model. Most of the computation can still be reduced through folded inference, and uni cross-attention only adds a small amount of inference cost to capture interest in the estimated item from the pre-trained model.

Furthermore, unlike the sequential recommendation tasks that rely solely on a single CLS token or a ``[mask]'' token for prediction, discerning useful information from hundreds of tokens in user behavior for CTR prediction also poses a significant challenge.
In this paper, we develop a querying transformer encoder that uses a small number of learnable query tokens to aggregate hundreds of user behavior tokens before passing them to the CTR model. 
This encoder can further enhance the performance of downstream tasks.

To validate the effectiveness of our proposed SRP4CTR method, we conducted extensive experiments on different sequential recommendation pre-trained models. The results demonstrate that SRP4CTR can enhance the performance of CTR tasks based on various pre-trained models while maintaining low inference costs. 

The contributions of this paper are as follows: 1) We analyze the inference costs, demonstrating that integrating CTR prediction with sequential recommendation pre-training is practical for industrial recommendation systems. 2) We propose a new prediction framework, which can transfer information from the self-supervised pre-trained model at a low cost. 3) We conducted complete offline experiments and online validations to prove the effectiveness of our proposed method.

\section{METHODOLOGY}
\begin{figure}[!t]
\centering
	\includegraphics[width=0.95\linewidth]{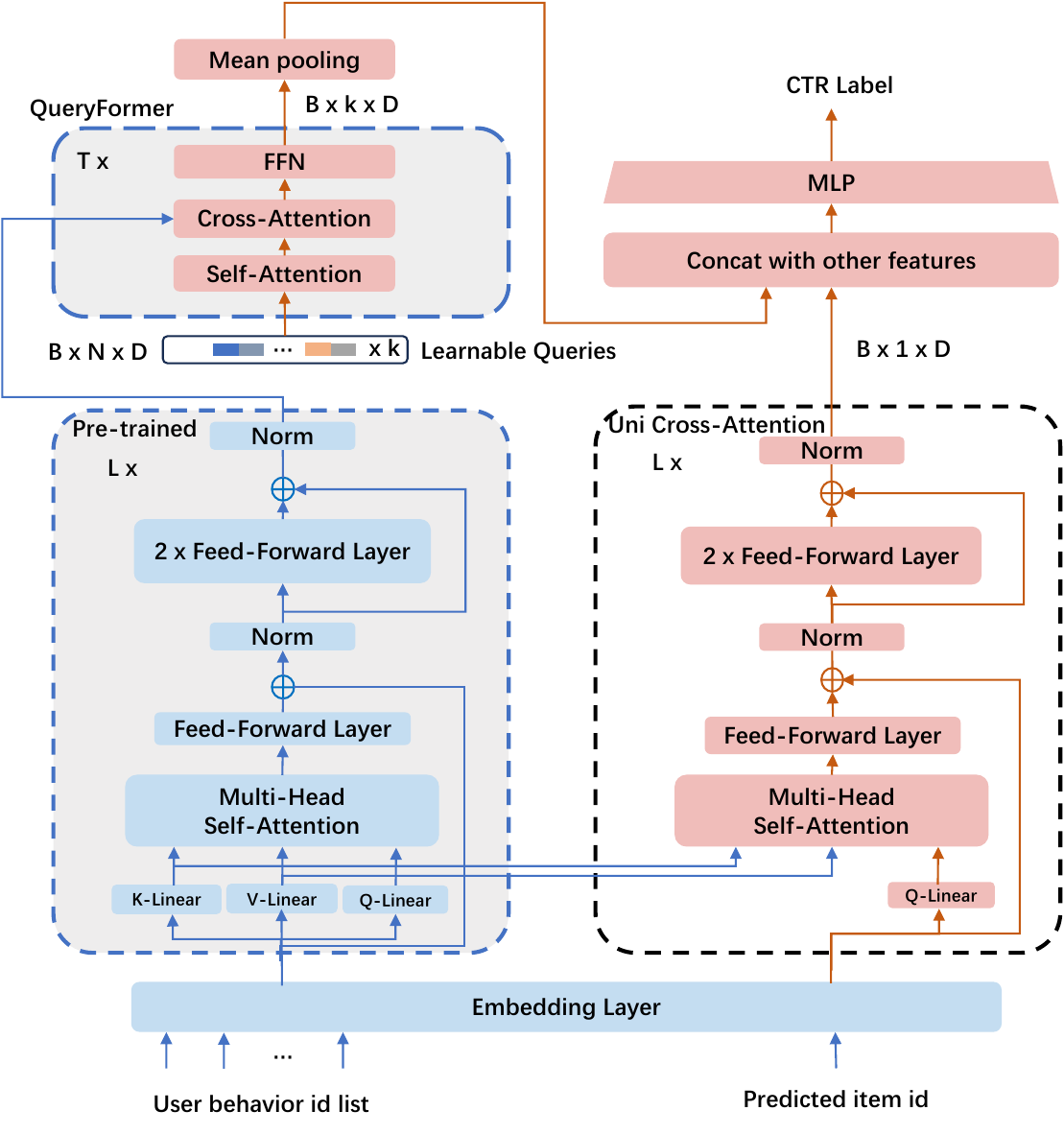}
	\vspace{-0.5em}
	\captionsetup{font={small}}
	
	\caption{Overall architecture of our SRP4CTR method. The gray area represents the part that can be accelerated through folded inference. }
	\vspace{-1.0em}
    \label{fig:overall}
   
\end{figure}

SRP4CTR is designed to leverage transformer-based sequential recommendation pre-trained models to enhance the learning effectiveness of downstream CTR tasks. The overview of SRP4CTR is presented in Figure ~\ref{fig:overall}. In the following sections, we first introduce the base pre-trained sequential recommendation model with side information. Then, we will describe how we transfer knowledge from the pre-trained model for estimated items in CTR prediction.

\subsection{Sequential Recommendation Pre-training}
Self-supervised pre-training is used to tackle data sparsity issues by enhancing data representation \cite{zhou2020s3}, and improve recommendation performance for long-tail items \cite{yao2021self}. 
Following the approach of \cite{sun2019bert4rec}, we utilize a bidirectional transformer for modeling and a Cloze objective for self-supervised pre-training. 
However, unlike most sequence recommendation tasks that only require modeling the item ID, real-world recommendation systems involve user click logs that contain not just the item ID but also a variety of side information. For a pre-trained model whose downstream task is CTR prediction, it is essential to encode not only the item ID but also the corresponding side information. To address this, we propose the Fine-Grained BERT (FG-BERT) to accomplish the task.

For generalization, given a sequence of user-item interaction record with length $L$, $i$-th elements $\mathbf{S}_i$ is consisted of M corresponding item-related features $f_{i1}, ..., f_{iM}$ such as item ID, tags, price, etc., and N behavior actions $b_{i1}, ..., f_{iN}$ such as types of interaction, count of click, etc., formally: $\mathbf{S}_i = \left[(f_{i1}, f_{i2}, ..., f_{iM}), (b_{i1}, b_{i2},...,b_{iN})\right]$.

\begin{figure}[!t]

\centering
	\includegraphics[width=0.80\linewidth]{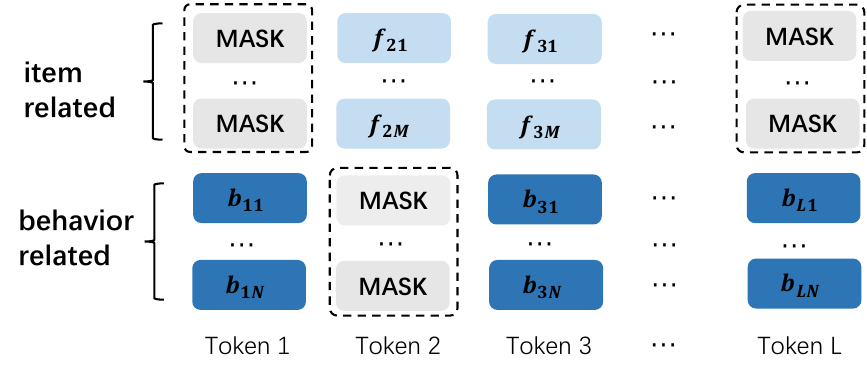}
	\vspace{-0.5em}
	\captionsetup{font={small}}
	
	\caption{Multi-attribute mask in Fine-Grained BERT.}
	\vspace{-1.0em}
    \label{fig:fgbert}
\end{figure}

FG-BERT enhances the encoding effectiveness of BERT by further introducing a training objective that predicts behavior-related features. 
Specifically, since the item ID and its corresponding attribute features are explicitly associated, we directly sum the embeddings of item-related features to obtain an item generalized representation $\mathbf{x}_i$ of $i$-th record. 
Similarly, we sum the embeddings of all behavior-related features to obtain the side info representation ${\mathbf{s}_i}$. 
As shown in Figure ~\ref{fig:fgbert}, in the FG-BERT, we introduce two different types of random mask: the item-related mask and the behavior-related mask.
For behavior-related masks, we remove all associated behavior actions for one element. The mask is denoted as $\pmb{s^*}=\{\mathbf{s}_{j_1}, \mathbf{s}_{j_2}, \cdots, \mathbf{s}_{j_p}\}, j_p < L$, with the remaining set denoted as $\tilde{\pmb{s}}$. 
Correspondingly, the set of item-related masks is denoted as $\pmb{x^*}=\{\mathbf{x}_{k_1}, \mathbf{x}_{k_2}, \cdots, \mathbf{x}_{k_q}\}, k_q < L$, with the remaining set denoted as $\tilde{\pmb{x}}$. For FG-BERT, the probability of $p(\pmb{s^*})$ and $p(\pmb{x^*})$ is given as:

\begin{equation}
\begin{aligned}
    p(\pmb{x}^*;\Theta)&=\prod_{i}^{q} p(\mathbf{x}_{k_i} | \tilde{\pmb{x}}, \tilde{\pmb{s}};\Theta) \\
    p(\pmb{s}^*;\Theta)&=\prod_{i}^p \prod_{n}^N p(b_{j_in}| \tilde{\pmb{x}}, \tilde{\pmb{s}};\Theta) \\
\end{aligned}    
\end{equation}
where $\Theta$ is the learnable parameter of the model. To maximize the probabilities $p(\pmb{x}^*;\Theta)$ and $p(\pmb{s}^*;\Theta)$, FG-BERT employs a multi-task approach, optimizing the aforementioned objectives with cross-entropy (CE) loss.




\subsection{Fine-tune for CTR Prediction}
\subsubsection{uni cross-attention block} For downstream CTR tasks, we introduce a uni cross-attention that enables the predicted item to transfer corresponding information from the pre-trained model. 
As illustrated in Figure ~\ref{fig:overall}, in uni cross-attention, the queries are composed of predicted items, while Key and Value are derived from the user behavior sequence representation at the same layer. 
Although the predicted item shares the same representational information as the input of the pre-trained model, the attention mechanism learned under self-supervised differs significantly from the attention mechanism required for CTR tasks. 
Therefore, we untie the parameter sharing between the uni cross-attention block and the pre-trained model, only sharing the input's embedding parameters and the projection parameters for Key and Value at each layer. 
After multi-head self-attention, the encoded feature of the predicted item is passed through the feed-forward network and residual networks, as commonly processed in the transformer-based layer.
Besides, a learnable variable is set as position embedding about the predicted item to circumvent issues with changes in positional semantics. 

During the deployment phase, the uni cross-attention block only adds a minimal amount of extra computation. The primary computational load, carried by the pre-trained model, can still be accelerated using folded inference.

\subsubsection{Querying Transformer}
Though encoding from the pre-trained model, the features generated may not align perfectly with the information required for CTR tasks, because the goal of pre-training mismatches with CTR training objective.
For better alignment, as shown in Figure ~\ref{fig:overall}, we introduce a novel Querying Transformer that employs $K$ learnable queries as input (where $K < L$) and encoded by cross-attention with the output of the pre-trained model as Key and Value. 
This module is expected to achieve discrimination of user interests. 
Additionally, in practical scenarios, we also use user or context features, such as gender, age, or hour, as the initialization for queries, which are then mapped into multiple different queries to accommodate the diverse interest distributions of different users.
Finally, the Querying Transformer only processes user information, which enables it to be folded inference.

\section{EXPERIMENT}
\subsection{Offline Experiments}
\subsubsection{Experimental Setups}
We conducted experiments using two offline public datasets: MovieLens-20M\cite{harper2015movielens} and Taobao\cite{zhou2018deep}. 
In the case of MovieLens, similarly to \cite{zhou2018deep}, the task is to predict whether a user is likely to assign a rating exceeding 3(denoted as a positive label). 
We leverage features derived from users' historical interactions to evaluate the most recent 10 interactions. 
We employ movie\_id and movie\_cate\_id as item-related features and the rating score from the log as the behavior-related feature.
For the Taobao dataset, we use the item cate\_id and brand\_id as item-related features, and the behavioral tag serves as the behavior-related feature.
Additionally, we have executed a data sampling in the Taobao dataset, selecting the top 50\% of users with the most behaviors.

In this work, we utilize the Adam optimizer and apply polynomial decay to the learning rate. 
During the pre-training phase, we use the same architecture as in \cite{sun2019bert4rec} consisting of a two-layer transformer with the maximum sequence length $L = 200$. For the MovieLens dataset, we train with a batch size of 512 for 100,000 steps. 
For the Taobao dataset, due to GPU memory constraints, we train with a batch size of 256 for 200,000 steps. 
During the fine-tuning phase, with Taobao's token count reaching 500,000 compared to MovieLens's token count of only 20,000, there is a higher risk of overfitting for MovieLens. Consequently, all parameters in the pre-trained model are learnable in the Taobao dataset but frozen in the MovieLens.
We train for 10 epochs and select the most effective checkpoint as the benchmark for performance evaluation.
Additionally, we use Area Under the Curve(AUC) as a metric to evaluate the performance of the CTR task.

\subsubsection{Overall Performance}
The overall performance of SRP4CTR is presented in Table ~\ref{tab:compare}. We primarily compared five traditional CTR-based methods: PNN\cite{qu2016product}, BST\cite{chen2019behavior},DIN\cite{zhou2018deep}, DIEN\cite{zhou2019deep}, and CAN\cite{bian2022can}. 
Our approach demonstrated notable improvements over conventional end-to-end modeling methods among various datasets. 

Furthermore, we explored the impact of different pre-training methods on SRP4CTR, including the original BERT\cite{sun2019bert4rec}, ELECTRA\cite{clark2020electra}, S3\cite{zhou2020s3}, and our proposed FG-BERT. 
Each experimental group included two models. We compared the performance of MP\cite{yuan2020parameter} and SRP4CTR on different pre-trained models. 
MP is a simple method of transferring a pre-trained model to downstream tasks. In the fine-tuning phase, this method inserts a learnable model patch into the pre-trained model while the rest of the parameters are frozen. 
Subsequently, the representation produced by the pre-trained model is concatenated with the embedding of the estimated item for prediction. 
MP enables a more direct comparison of the effectiveness of different pre-training methods.
From Table 1, our proposed FG-BERT outperforms other pre-training approaches. Furthermore, on the MovieLens dataset, the pre-trained model can surpass CTR-based methods solely through MP. However, for the sparser Taobao dataset, MP underperforms compared to CTR-based models.
\subsubsection{Long Tail Performance}
In Table ~\ref{tab:longtail}, we further explored the effectiveness of SRP4CTR for recommending long-tail items. 
We defined the bottom 20\% of items in terms of their occurrence in the dataset as long-tail items and compared the performance of our method and traditional CTR prediction methods on this sub-dataset. 
Traditional CTR prediction methods like DIN and CAN tend to learn more effectively from hot items, resulting in a significant discrepancy between the AUC for long-tail items and the average AUC across all items. 
However, through pre-training, both the MP and SRP4CTR can greatly narrow the learning gap between long-tail items and other items.

\begin{table}[!t]
\captionsetup{font={small}}
\caption{Results (AUC) on public datasets.}
\scalebox{1.00}{
\begin{tabular}{cl|c|c}

		\toprule
  \multicolumn{2}{c|}{\textbf{Methods}}  & \multicolumn{1}{c|}{\textbf{MovieLens}}  & \multicolumn{1}{c}{\textbf{Taobao}}  \\
          \hline
\multirow{5}{*}{CTR based} & PNN\cite{qu2016product} & 0.7376 & 0.6006 \\
\multirow{5}{*}{} & BST\cite{chen2019behavior} & 0.7359 & 0.6021 \\
\multirow{5}{*}{} & DIN\cite{zhou2018deep} & 0.7466 & 0.6066  \\
\multirow{5}{*}{} & DIEN\cite{zhou2019deep} & 0.7494  & 0.6178 \\
\multirow{5}{*}{} & CAN\cite{bian2022can} & 0.7507  & 0.6018  \\
\hline
\multirow{2}{*}{BERT\cite{sun2019bert4rec}} & BERT-MP & 0.7595 & 0.5633  \\
\multirow{2}{*}{} & \textbf{BERT-SRP4CTR} & 0.7696 & 0.6208 \\
\hline
\multirow{2}{*}{ELECTRA\cite{clark2020electra}} & ELECTRA-MP & 0.7559  & 0.5580   \\
\multirow{2}{*}{} & \textbf{ELECTRA-SPR4CTR} & 0.7603  & 0.6197  \\
\hline
\multirow{2}{*}{S3\cite{zhou2020s3}} & S3-MP & 0.7526 & 0.5576  \\
\multirow{2}{*}{} & \textbf{S3-SRP4CTR} & 0.7592 & 0.6203  \\
\hline
\multirow{2}{*}{FG-BERT} & FG-BERT-MP & 0.7721  &0.5651   \\
\multirow{2}{*}{} & \textbf{SPR4CTR} & \textbf{0.7817} & \textbf{0.6230} \\
\hline

\end{tabular}
}
\label{tab:compare}
\end{table}

\begin{table}[!t]
\captionsetup{font={small}}
\caption{Recommendation performance of long-tail items 
. The diff indicator represents the difference in AUC between long-tail products and the AUC of all products.
}
\scalebox{1.00}{
\begin{tabular}{l|cc|cc}
\toprule
\multicolumn{1}{l|}{\multirow{2}{*}{\textbf{Methods}}}
  & \multicolumn{2}{c|}{\textbf{MovieLens}}  & \multicolumn{2}{c}{\textbf{Taobao}}  \\
  \cline{2-5}
 & \textbf{AUC} & \textbf{AUC diff} & \textbf{AUC} & \textbf{AUC diff} \\ 
  \hline
DIN & 0.7205 & -0.0261 & 0.6044 & -0.0022 \\
CAN & 0.7208 & -0.0299 & 0.5957 & -0.0061 \\
FG-BERT-MP & 0.7591 & \textbf{-0.0130} & 0.5686 & \textbf{+0.0035} \\
\textbf{SRP4CTR} & \textbf{0.7679} & -0.0139 & \textbf{0.6225} & -0.0005 \\
\hline

\end{tabular}
}
\label{tab:longtail}
\end{table}

\subsubsection{Ablation Studies}
We conducted detailed ablation studies on SRP4CTR, with the results presented in Table ~\ref{tab:ablation}. 
The comparisons include our proposed querying transformer(qFormer) and uni cross-attention module, as well as the impact of other training paradigms like scratch. Scratch denotes training SRP4CTR directly in an end-to-end manner.
Moreover, for the uni cross-attention, we also examined the impact of parameter tying.  The results validate the effectiveness of our proposed methods.

\begin{table}[!t]
\captionsetup{font={small}}
\caption{Ablation studies of our SRP4CTR method.(AUC)}
\scalebox{1.00}{
\begin{tabular}{l|c|c}

		\toprule
 \textbf{Methods} & \multicolumn{1}{c|}{\textbf{MovieLens}}  & \multicolumn{1}{c}{\textbf{Taobao}}  \\
          \hline
SRP4CTR-Scratch & 0.7740 & 0.6107 \\
SRP4CTR w/o uni-att \& qFormer & 0.7651 & 0.6099 \\
SRP4CTR w/o uni-att & 0.7726 & 0.6116 \\
SRP4CTR w/o qFormer & 0.7808 & 0.6220  \\
SRP4CTR w tying uni-att & 0.7728 & 0.6153 \\
\textbf{SRP4CTR} & \textbf{0.7817} & \textbf{0.6230} \\
\hline

\end{tabular}
}
\label{tab:ablation}
\end{table}
\subsubsection{Inference Cost Analysis}
In Table ~\ref{tab:inference}, we have conducted a comparison of the inference costs between SRP4CTR and other methodologies.  We propose a new metric \textit{efficiency-FLOPs}, which denotes the FLOPs when the inference batch size is set to one. This metric is indicative of the computational complexity that truly impacts the precision of the model within a folded inference framework. Furthermore, \textit{inference-FLOPs} represents the computational complexity for a single batch when inference is carried out using folded inference with a batch size of 100 \footnote{Our online deployment performs inference with a batch size of 100.}, thereby reflecting the cost of the model's inference. 

As the original implementations of DIN and CAN did not take into account inference performance, we have introduced two variants, DIN$\dagger$ and CAN$\dagger$. 
These variants can be accelerated by folded inference, designed to align as closely as possible with the original versions in terms of efficiency-FLOPs.  From Table ~\ref{tab:inference}, SRP4CTR is capable of delivering an increase in efficiency-FLOPs by more than double in comparison to DIN$\dagger$ and CAN$\dagger$, with only a slight rise in inference-FLOPs.

\begin{table}[!t]
\captionsetup{font={small}}
\caption{Inference cost (FLOPs(M)) of different models.
}
\scalebox{0.90}{
\begin{tabular}{l|c|c|c|c|c}
	\toprule
\textbf{Metrics} & DIN & DIN$\dagger$ & CAN & CAN$\dagger$ & \textbf{SRP4CTR} \\
		\hline
 
efficiency-FLOPs & 10.01 & 8.96 & 10.12 & 8.99 & \textbf{26.88} \\
\hline
inference-FLOPs & 989.59 & 51.22 & 1002.21 & 59.64 & \textbf{64.56} \\
\hline
\makecell{inference-FLOPs/ \\ efficiency-FLOPs} & 98.86 & 5.72 & 99.03 & 6.63 & \textbf{2.40} \\
\hline
\end{tabular}
}
\label{tab:inference}
\end{table}

\subsection{Online A/B test}
We deploy our method to the Meituan Takeaway recommender system, one of the largest takeaway platforms in China. We have conducted a one-month online A/B test since June 2023. The SRP4CTR increased the GMV(Gross Merchandise Volume) by 1.66\% and the CTR by 0.70\% in our main recommendation scenarios.  
From the perspective of efficiency, compared with the previous state-of-art baseline(DIN+MMOE\cite{ma2018modeling}), our model brings an 182\% increase in efficiency-FLOPs. Meanwhile, the inference-FLOPs only increase by 21\% with the folded inference framework.

\section{Conclusions}
In this paper, we propose a new method named SRP4CTR to improve CTR prediction by adapting the sequential recommendation pre-trained model at a low cost. Concretely, we first analyze the inference costs associated with introducing pre-trained models into CTR tasks. Then, we introduce a Fine-Grained BERT and uni cross-attention mechanism to accomplish the knowledge transfer. Our method has achieved significant growth in Gross Merchandise Volume and has been deployed in Meituan Waimai recommendation scenarios since July 2023.

\bibliographystyle{ACM-Reference-Format}
\bibliography{sample-base}


\begin{thebibliography}{22}


\ifx \showCODEN    \undefined \def \showCODEN     #1{\unskip}     \fi
\ifx \showDOI      \undefined \def \showDOI       #1{#1}\fi
\ifx \showISBNx    \undefined \def \showISBNx     #1{\unskip}     \fi
\ifx \showISBNxiii \undefined \def \showISBNxiii  #1{\unskip}     \fi
\ifx \showISSN     \undefined \def \showISSN      #1{\unskip}     \fi
\ifx \showLCCN     \undefined \def \showLCCN      #1{\unskip}     \fi
\ifx \shownote     \undefined \def \shownote      #1{#1}          \fi
\ifx \showarticletitle \undefined \def \showarticletitle #1{#1}   \fi
\ifx \showURL      \undefined \def \showURL       {\relax}        \fi
\providecommand\bibfield[2]{#2}
\providecommand\bibinfo[2]{#2}
\providecommand\natexlab[1]{#1}
\providecommand\showeprint[2][]{arXiv:#2}

\bibitem[Bian et~al\mbox{.}(2022)]%
        {bian2022can}
\bibfield{author}{\bibinfo{person}{Weijie Bian}, \bibinfo{person}{Kailun Wu}, \bibinfo{person}{Lejian Ren}, \bibinfo{person}{Qi Pi}, \bibinfo{person}{Yujing Zhang}, \bibinfo{person}{Can Xiao}, \bibinfo{person}{Xiang-Rong Sheng}, \bibinfo{person}{Yong-Nan Zhu}, \bibinfo{person}{Zhangming Chan}, \bibinfo{person}{Na Mou}, {et~al\mbox{.}}} \bibinfo{year}{2022}\natexlab{}.
\newblock \showarticletitle{CAN: feature co-action network for click-through rate prediction}. In \bibinfo{booktitle}{\emph{Proceedings of the fifteenth ACM international conference on web search and data mining}}. \bibinfo{pages}{57--65}.
\newblock


\bibitem[Chang et~al\mbox{.}(2023)]%
        {chang2023twin}
\bibfield{author}{\bibinfo{person}{Jianxin Chang}, \bibinfo{person}{Chenbin Zhang}, \bibinfo{person}{Zhiyi Fu}, \bibinfo{person}{Xiaoxue Zang}, \bibinfo{person}{Lin Guan}, \bibinfo{person}{Jing Lu}, \bibinfo{person}{Yiqun Hui}, \bibinfo{person}{Dewei Leng}, \bibinfo{person}{Yanan Niu}, \bibinfo{person}{Yang Song}, {et~al\mbox{.}}} \bibinfo{year}{2023}\natexlab{}.
\newblock \showarticletitle{TWIN: TWo-stage Interest Network for Lifelong User Behavior Modeling in CTR Prediction at Kuaishou}.
\newblock \bibinfo{journal}{\emph{arXiv preprint arXiv:2302.02352}} (\bibinfo{year}{2023}).
\newblock


\bibitem[Chen et~al\mbox{.}(2019)]%
        {chen2019behavior}
\bibfield{author}{\bibinfo{person}{Qiwei Chen}, \bibinfo{person}{Huan Zhao}, \bibinfo{person}{Wei Li}, \bibinfo{person}{Pipei Huang}, {and} \bibinfo{person}{Wenwu Ou}.} \bibinfo{year}{2019}\natexlab{}.
\newblock \showarticletitle{Behavior sequence transformer for e-commerce recommendation in alibaba}. In \bibinfo{booktitle}{\emph{Proceedings of the 1st international workshop on deep learning practice for high-dimensional sparse data}}. \bibinfo{pages}{1--4}.
\newblock


\bibitem[Chitlangia et~al\mbox{.}(2023)]%
        {chitlangia2023scaling}
\bibfield{author}{\bibinfo{person}{Sharad Chitlangia}, \bibinfo{person}{Krishna~Reddy Kesari}, {and} \bibinfo{person}{Rajat Agarwal}.} \bibinfo{year}{2023}\natexlab{}.
\newblock \showarticletitle{Scaling generative pre-training for user ad activity sequences}.
\newblock  (\bibinfo{year}{2023}).
\newblock


\bibitem[Clark et~al\mbox{.}(2020)]%
        {clark2020electra}
\bibfield{author}{\bibinfo{person}{Kevin Clark}, \bibinfo{person}{Minh-Thang Luong}, \bibinfo{person}{Quoc~V Le}, {and} \bibinfo{person}{Christopher~D Manning}.} \bibinfo{year}{2020}\natexlab{}.
\newblock \showarticletitle{Electra: Pre-training text encoders as discriminators rather than generators}.
\newblock \bibinfo{journal}{\emph{arXiv preprint arXiv:2003.10555}} (\bibinfo{year}{2020}).
\newblock


\bibitem[de~Souza Pereira~Moreira et~al\mbox{.}(2021)]%
        {de2021transformers4rec}
\bibfield{author}{\bibinfo{person}{Gabriel de Souza Pereira~Moreira}, \bibinfo{person}{Sara Rabhi}, \bibinfo{person}{Jeong~Min Lee}, \bibinfo{person}{Ronay Ak}, {and} \bibinfo{person}{Even Oldridge}.} \bibinfo{year}{2021}\natexlab{}.
\newblock \showarticletitle{Transformers4rec: Bridging the gap between nlp and sequential/session-based recommendation}. In \bibinfo{booktitle}{\emph{Proceedings of the 15th ACM Conference on Recommender Systems}}. \bibinfo{pages}{143--153}.
\newblock


\bibitem[Harper and Konstan(2015)]%
        {harper2015movielens}
\bibfield{author}{\bibinfo{person}{F~Maxwell Harper} {and} \bibinfo{person}{Joseph~A Konstan}.} \bibinfo{year}{2015}\natexlab{}.
\newblock \showarticletitle{The movielens datasets: History and context}.
\newblock \bibinfo{journal}{\emph{Acm transactions on interactive intelligent systems (tiis)}} \bibinfo{volume}{5}, \bibinfo{number}{4} (\bibinfo{year}{2015}), \bibinfo{pages}{1--19}.
\newblock


\bibitem[Liao(2020)]%
        {liao2020effectiveness}
\bibfield{author}{\bibinfo{person}{Yiping Liao}.} \bibinfo{year}{2020}\natexlab{}.
\newblock \bibinfo{title}{On the Effectiveness of Self-supervised Pre-training for Modeling User Behavior Sequences}.
\newblock
\newblock


\bibitem[Lin et~al\mbox{.}(2023)]%
        {lin2023map}
\bibfield{author}{\bibinfo{person}{Jianghao Lin}, \bibinfo{person}{Yanru Qu}, \bibinfo{person}{Wei Guo}, \bibinfo{person}{Xinyi Dai}, \bibinfo{person}{Ruiming Tang}, \bibinfo{person}{Yong Yu}, {and} \bibinfo{person}{Weinan Zhang}.} \bibinfo{year}{2023}\natexlab{}.
\newblock \showarticletitle{MAP: A Model-agnostic Pretraining Framework for Click-through Rate Prediction}. In \bibinfo{booktitle}{\emph{Proceedings of the 29th ACM SIGKDD Conference on Knowledge Discovery and Data Mining}}. \bibinfo{pages}{1384--1395}.
\newblock


\bibitem[Liu et~al\mbox{.}(2021)]%
        {liu2021noninvasive}
\bibfield{author}{\bibinfo{person}{Chang Liu}, \bibinfo{person}{Xiaoguang Li}, \bibinfo{person}{Guohao Cai}, \bibinfo{person}{Zhenhua Dong}, \bibinfo{person}{Hong Zhu}, {and} \bibinfo{person}{Lifeng Shang}.} \bibinfo{year}{2021}\natexlab{}.
\newblock \showarticletitle{Noninvasive self-attention for side information fusion in sequential recommendation}. In \bibinfo{booktitle}{\emph{Proceedings of the AAAI Conference on Artificial Intelligence}}, Vol.~\bibinfo{volume}{35}. \bibinfo{pages}{4249--4256}.
\newblock


\bibitem[Liu et~al\mbox{.}(2022)]%
        {liu2022boosting}
\bibfield{author}{\bibinfo{person}{Qijiong Liu}, \bibinfo{person}{Jieming Zhu}, \bibinfo{person}{Quanyu Dai}, {and} \bibinfo{person}{Xiao-Ming Wu}.} \bibinfo{year}{2022}\natexlab{}.
\newblock \showarticletitle{Boosting deep CTR prediction with a plug-and-play pre-trainer for news recommendation}. In \bibinfo{booktitle}{\emph{Proceedings of the 29th International Conference on Computational Linguistics}}. \bibinfo{pages}{2823--2833}.
\newblock


\bibitem[Ma et~al\mbox{.}(2018)]%
        {ma2018modeling}
\bibfield{author}{\bibinfo{person}{Jiaqi Ma}, \bibinfo{person}{Zhe Zhao}, \bibinfo{person}{Xinyang Yi}, \bibinfo{person}{Jilin Chen}, \bibinfo{person}{Lichan Hong}, {and} \bibinfo{person}{Ed~H Chi}.} \bibinfo{year}{2018}\natexlab{}.
\newblock \showarticletitle{Modeling task relationships in multi-task learning with multi-gate mixture-of-experts}. In \bibinfo{booktitle}{\emph{Proceedings of the 24th ACM SIGKDD international conference on knowledge discovery \& data mining}}. \bibinfo{pages}{1930--1939}.
\newblock


\bibitem[Pi et~al\mbox{.}(2020)]%
        {pi2020search}
\bibfield{author}{\bibinfo{person}{Qi Pi}, \bibinfo{person}{Guorui Zhou}, \bibinfo{person}{Yujing Zhang}, \bibinfo{person}{Zhe Wang}, \bibinfo{person}{Lejian Ren}, \bibinfo{person}{Ying Fan}, \bibinfo{person}{Xiaoqiang Zhu}, {and} \bibinfo{person}{Kun Gai}.} \bibinfo{year}{2020}\natexlab{}.
\newblock \showarticletitle{Search-based user interest modeling with lifelong sequential behavior data for click-through rate prediction}. In \bibinfo{booktitle}{\emph{Proceedings of the 29th ACM International Conference on Information \& Knowledge Management}}. \bibinfo{pages}{2685--2692}.
\newblock


\bibitem[Qu et~al\mbox{.}(2016)]%
        {qu2016product}
\bibfield{author}{\bibinfo{person}{Yanru Qu}, \bibinfo{person}{Han Cai}, \bibinfo{person}{Kan Ren}, \bibinfo{person}{Weinan Zhang}, \bibinfo{person}{Yong Yu}, \bibinfo{person}{Ying Wen}, {and} \bibinfo{person}{Jun Wang}.} \bibinfo{year}{2016}\natexlab{}.
\newblock \showarticletitle{Product-based neural networks for user response prediction}. In \bibinfo{booktitle}{\emph{2016 IEEE 16th international conference on data mining (ICDM)}}. IEEE, \bibinfo{pages}{1149--1154}.
\newblock


\bibitem[Sun et~al\mbox{.}(2019)]%
        {sun2019bert4rec}
\bibfield{author}{\bibinfo{person}{Fei Sun}, \bibinfo{person}{Jun Liu}, \bibinfo{person}{Jian Wu}, \bibinfo{person}{Changhua Pei}, \bibinfo{person}{Xiao Lin}, \bibinfo{person}{Wenwu Ou}, {and} \bibinfo{person}{Peng Jiang}.} \bibinfo{year}{2019}\natexlab{}.
\newblock \showarticletitle{BERT4Rec: Sequential recommendation with bidirectional encoder representations from transformer}. In \bibinfo{booktitle}{\emph{Proceedings of the 28th ACM international conference on information and knowledge management}}. \bibinfo{pages}{1441--1450}.
\newblock


\bibitem[Wang et~al\mbox{.}(2023)]%
        {wang2023bert4ctr}
\bibfield{author}{\bibinfo{person}{Dong Wang}, \bibinfo{person}{Kav{\'e} Salamatian}, \bibinfo{person}{Yunqing Xia}, \bibinfo{person}{Weiwei Deng}, {and} \bibinfo{person}{Qi Zhang}.} \bibinfo{year}{2023}\natexlab{}.
\newblock \showarticletitle{BERT4CTR: An Efficient Framework to Combine Pre-trained Language Model with Non-textual Features for CTR Prediction}. In \bibinfo{booktitle}{\emph{Proceedings of the 29th ACM SIGKDD Conference on Knowledge Discovery and Data Mining}}. \bibinfo{pages}{5039--5050}.
\newblock


\bibitem[Yao et~al\mbox{.}(2021)]%
        {yao2021self}
\bibfield{author}{\bibinfo{person}{Tiansheng Yao}, \bibinfo{person}{Xinyang Yi}, \bibinfo{person}{Derek~Zhiyuan Cheng}, \bibinfo{person}{Felix Yu}, \bibinfo{person}{Ting Chen}, \bibinfo{person}{Aditya Menon}, \bibinfo{person}{Lichan Hong}, \bibinfo{person}{Ed~H Chi}, \bibinfo{person}{Steve Tjoa}, \bibinfo{person}{Jieqi Kang}, {et~al\mbox{.}}} \bibinfo{year}{2021}\natexlab{}.
\newblock \showarticletitle{Self-supervised learning for large-scale item recommendations}. In \bibinfo{booktitle}{\emph{Proceedings of the 30th ACM International Conference on Information \& Knowledge Management}}. \bibinfo{pages}{4321--4330}.
\newblock


\bibitem[Yuan et~al\mbox{.}(2020)]%
        {yuan2020parameter}
\bibfield{author}{\bibinfo{person}{Fajie Yuan}, \bibinfo{person}{Xiangnan He}, \bibinfo{person}{Alexandros Karatzoglou}, {and} \bibinfo{person}{Liguang Zhang}.} \bibinfo{year}{2020}\natexlab{}.
\newblock \showarticletitle{Parameter-efficient transfer from sequential behaviors for user modeling and recommendation}. In \bibinfo{booktitle}{\emph{Proceedings of the 43rd International ACM SIGIR conference on research and development in Information Retrieval}}. \bibinfo{pages}{1469--1478}.
\newblock


\bibitem[Zhai et~al\mbox{.}(2024)]%
        {zhai2024actions}
\bibfield{author}{\bibinfo{person}{Jiaqi Zhai}, \bibinfo{person}{Lucy Liao}, \bibinfo{person}{Xing Liu}, \bibinfo{person}{Yueming Wang}, \bibinfo{person}{Rui Li}, \bibinfo{person}{Xuan Cao}, \bibinfo{person}{Leon Gao}, \bibinfo{person}{Zhaojie Gong}, \bibinfo{person}{Fangda Gu}, \bibinfo{person}{Michael He}, {et~al\mbox{.}}} \bibinfo{year}{2024}\natexlab{}.
\newblock \showarticletitle{Actions Speak Louder than Words: Trillion-Parameter Sequential Transducers for Generative Recommendations}.
\newblock \bibinfo{journal}{\emph{arXiv preprint arXiv:2402.17152}} (\bibinfo{year}{2024}).
\newblock


\bibitem[Zhou et~al\mbox{.}(2019)]%
        {zhou2019deep}
\bibfield{author}{\bibinfo{person}{Guorui Zhou}, \bibinfo{person}{Na Mou}, \bibinfo{person}{Ying Fan}, \bibinfo{person}{Qi Pi}, \bibinfo{person}{Weijie Bian}, \bibinfo{person}{Chang Zhou}, \bibinfo{person}{Xiaoqiang Zhu}, {and} \bibinfo{person}{Kun Gai}.} \bibinfo{year}{2019}\natexlab{}.
\newblock \showarticletitle{Deep interest evolution network for click-through rate prediction}. In \bibinfo{booktitle}{\emph{Proceedings of the AAAI conference on artificial intelligence}}, Vol.~\bibinfo{volume}{33}. \bibinfo{pages}{5941--5948}.
\newblock


\bibitem[Zhou et~al\mbox{.}(2018)]%
        {zhou2018deep}
\bibfield{author}{\bibinfo{person}{Guorui Zhou}, \bibinfo{person}{Xiaoqiang Zhu}, \bibinfo{person}{Chenru Song}, \bibinfo{person}{Ying Fan}, \bibinfo{person}{Han Zhu}, \bibinfo{person}{Xiao Ma}, \bibinfo{person}{Yanghui Yan}, \bibinfo{person}{Junqi Jin}, \bibinfo{person}{Han Li}, {and} \bibinfo{person}{Kun Gai}.} \bibinfo{year}{2018}\natexlab{}.
\newblock \showarticletitle{Deep interest network for click-through rate prediction}. In \bibinfo{booktitle}{\emph{Proceedings of the 24th ACM SIGKDD international conference on knowledge discovery \& data mining}}. \bibinfo{pages}{1059--1068}.
\newblock


\bibitem[Zhou et~al\mbox{.}(2020)]%
        {zhou2020s3}
\bibfield{author}{\bibinfo{person}{Kun Zhou}, \bibinfo{person}{Hui Wang}, \bibinfo{person}{Wayne~Xin Zhao}, \bibinfo{person}{Yutao Zhu}, \bibinfo{person}{Sirui Wang}, \bibinfo{person}{Fuzheng Zhang}, \bibinfo{person}{Zhongyuan Wang}, {and} \bibinfo{person}{Ji-Rong Wen}.} \bibinfo{year}{2020}\natexlab{}.
\newblock \showarticletitle{S3-rec: Self-supervised learning for sequential recommendation with mutual information maximization}. In \bibinfo{booktitle}{\emph{Proceedings of the 29th ACM international conference on information \& knowledge management}}. \bibinfo{pages}{1893--1902}.
\newblock


\end{thebibliography}

\appendix

\end{document}